\documentclass[12pt,a4paper]{article} 

\usepackage{latexsym} 
\usepackage{amssymb} 
\usepackage{epsfig}    
\usepackage{stmaryrd}   
\newcommand{\beq}{\begin{equation}}
\newcommand{\eeq}{\end{equation}}
\newcommand{\bea}{\begin{eqnarray}}
\newcommand{\eea}{\end{eqnarray}}

\newcommand{\Tr}{\mbox{Tr}\,}

\def\slashchar#1{\setbox0=\hbox{$#1$}           % set a box for #1 
   \dimen0=\wd0                                 % and get its size
   \setbox1=\hbox{/} \dimen1=\wd1               % get size of /
   \ifdim\dimen0>\dimen1                        % #1 is bigger
      \rlap{\hbox to \dimen0{\hfil/\hfil}}      % so center / in box
      #1                                        % and print #1
   \else                                        % / is bigger
      \rlap{\hbox to \dimen1{\hfil$#1$\hfil}}   % so center #1
      /                                         % and print /
   \fi}

\begin{document}

%header%]]]
%title page%[[[
%title%[[[
\title{\bf The pressure of QED from the\\
two-loop 2PI effective action}
%title%]]]
%author%[[[
\author{
Sz. Bors\'anyi\thanks{email: s.borsanyi@sussex.ac.uk} $\,^{a}$ and 
U. Reinosa\thanks{email: reinosa@thphys.uni-heidelberg.de} $\,^{b}$,
\\[0.4cm]
\normalsize{$^a$ Department of Physics and Astronomy, University of Sussex,}\\[-0.1cm]
\normalsize{Brighton, East Sussex BN1 9QH, United Kingdom.}\\
\normalsize{$^b$ Institut f\"ur Theoretische Physik, Universit\"at Heidelberg,}\\[-0.1cm]
\normalsize{Philosophenweg 16, 69120 Heidelberg, Germany.}\\
}
%author%]]]
% date and latex stuff %[[[
\date{}
%\\[1ex] 
%\preprintno}

\begin{titlepage}
\maketitle
\def\thepage{}          % No page number on title page

% date and latex stuff %]]]
%abstract%[[[
\begin{abstract}
We compute the pressure of hot quantum electrodynamics from the two-loop
truncation of the 2PI effective action. Since the 2PI resummation guarantees
gauge-fixing independence only up to the order of the truncation, our result
for the pressure presents a gauge
dependent contribution of $\mathcal{O}(e^4)$. We numerically characterize the
credibility of this gauge-dependent calculation and find that the uncertainty
due to gauge parameter dependence is under control for $\xi\lesssim 1$.
Our calculation also suggests that the choice of Landau gauge may minimize gauge-dependent effects.
\end{abstract}
%abstract%]]]
\end{titlepage}
\renewcommand{\thepage}{\arabic{page}}
%title page%]]]
% Introduction %[[[
% 2PI in general %[[[

The diagrammatic approach to finite temperature quantum field theories
heavily relies on the convergence properties of the used expansion scheme.
To cure the poor performance of the perturbative loop expansion,
various resummation schemes have been invented \cite{resum}.
Among these, the loop expansion of the two-particle-irreducible (2PI) effective
action implements a ladder resummation, which respects thermodynamical
consistency and energy conservation \cite{Luttinger:1960ua:andothers,
Cornwall:1974vz}.  These features make the 2PI scheme attractive
for nonequilibrium field theory applications \cite{Berges:2004yj}.  
A prerequisite for a nonequilibrium method
to be credible is, however, its reliability in equilibrium.  There, it is
important to check the convergence of expansion series of the 2PI effective
action. To this aim we calculated the notoriously ill-behaved pressure in a
scalar context in Ref.~\cite{Berges:2004hn}, and found a monotonous dependence
on the coupling constant as well as a relatively small next-to-leading order
correction to the pressure value even at couplings of $\mathcal{O}(1)$.

% 2PI in general %]]]
% 2PI and gauge %[[[
In the framework of gauge theories, however, the implementation of this
approximation scheme suffers from various difficulties. One of these is that
thermodynamic observables computed within this scheme are gauge fixing
independent only up to the order of the truncation. One can illustrate this point by considering the issue of gauge parameter dependence in the covariant gauge. For vanishing background fields, the 2PI effective action is a functional of the fermion, gauge and ghost propagators (respectively denoted by $D$, $G$ and $G_{\rm gh}$) which also depends on the gauge-fixing parameter~$\xi$: $\Gamma_{\rm 2PI}[D,G,G_{\rm gh};\xi]$.  The thermal pressure of the system is obtained by evaluating
$\Gamma_{\rm 2PI}$ at its stationary point\footnote{The barred propagators denote the solution of the stationarity equations: $\smash{\delta \Gamma_{\rm 2PI}/\delta
D=0}$, $\smash{\delta\Gamma_{\rm 2PI}/\delta G=0}$ and
$\smash{\delta\Gamma_{\rm 2PI}/\delta G_{\rm gh}=0}$.} $\smash{D=\bar D}$, $\smash{G=\bar G}$, $\smash{G_{\rm gh}=\bar G_{\rm gh}}$, for a given temperature $T$, and by subtracting the same calculation at zero temperature:
\begin{equation}
\mathcal{P}=-\left.\frac{T}{V}\Gamma_{\rm 2PI}[\bar D,\bar G,\bar G_{\rm gh};\xi]\right|^T_{T=0}\,.
\label{eq:pressure}
\end{equation}
It is possible to show that the $\xi$-dependence of $\mathcal{P}$ uniquely comes from the explicit $\xi$-dependence of $\Gamma_{\rm 2PI}$ and that it disappears if, in Fourier space,
\begin{equation}\label{eq:BRST}
\sum_{\mu\nu}q_\mu q_\nu \bar G_{\mu\nu}(q)=\xi\,.
\end{equation}
This last equation is the BRST identity for the propagator of the exact
theory \cite{LeBellac}, which may break in a truncated resummation.
Indeed, within a given truncation of the 2PI effective action, the gauge
symmetry does not impose any constraint \cite{Reinosa:2007vi} on the two-point function $\bar G$ above the order of the truncation.\footnote{Below this order, the truncated two-point function $\bar G$ coincides with the exact one which fulfills the BRST identity.}
Thus beyond this order Eq.~(\ref{eq:BRST}) breaks
and $\xi$-dependent contributions to the pressure appear.\footnote{More precisely, if one truncates the 2PI effective
action at L-loop order, one expects gauge dependences to appear at order
$e^{2L}$ \cite{Arrizabalaga:2002hn}.
}

% 2pi and gauge %]]]
% approx self consistent %[[[

A certain number of strategies can be put forward in order to try to cope with
these inconvenient features. The first possibility is to introduce further
approximations, on top of the loop expansion. This is the case of the {\it
approximately self-consistent resummations} introduced in
Ref. \cite{Blaizot:1999ip}. Using this method, a gauge
independent determination of the entropy of QCD has been possible and shows a
good agreement with lattice results down to temperatures about 2.5 times the
transition temperature. There is however no general understanding on how to
systematize this approach and evaluate higher orders in a gauge independent
manner.

% approx self consistent %]]]
% Reparametrization %[[[

Another possibility is to stick to the loop expansion of the 2PI effective
action but play with the freedom in the choice of field representations. Indeed,
the exact theory is invariant under reparametrization of the fields and one
could exploit this feature in order to define a loop expansion obeying certain
properties.  This idea has been discussed in Ref.~\cite{Leupold:2006bp} where it has
been applied to the linear sigma model in order to define a
systematic loop expansion of the 2PI effective action fulfilling Goldstone's
theorem at any order of approximation. Unfortunately no field representation is
yet known in gauge theories which would ensure that the BRST identity
(\ref{eq:BRST}) is fulfilled. 

% Reparametrization %]]]
% gauge-fixing dependent 2PI ea %[[[

It is finally possible to isolate gauge independent terms in the expression of
the pressure by separating contributions from different perturbative orders.
This means a re-expansion of the propagators $\bar D$ and $\bar G$ in powers of the coupling.
The resulting modified resummation scheme did not show a substantial
improvement of convergence \cite{Andersen:2004re}.

A different point of view is based on the experience that the 2PI loop
expansion is known to have good convergence properties
\cite{Berges:2004hn,Blaizot:2005wr}. One can expect that contributions above
the order of accuracy, and in particular gauge dependences are under control,
at least in a large range of coupling values.  In this paper we explore this possibility and consider the two-loop truncation of the 2PI effective action using the standard parametrization of the fields. We work in the covariant gauge with arbitrary gauge-fixing parameter $\xi$, which allows us to study how large gauge dependent contributions can be.

% gauge-fixing dependence of the 2PI ea %]]]
% aspect of renormalization%[[[

Before embarking on a numerical evaluation, one has however to pay special
attention to a second aspect, namely that of renormalization. The difficulty is
related to the fact that truncations of the 2PI effective action only resum
particular subclasses of perturbative diagrams for which (perturbative)
theorems do not apply. Recently a large effort has been put into extending
renormalization theorems to the particular classes of diagrams resummed by the
loop expansion of the 2PI effective action. This has been first achieved in the
framework of scalar theories \cite{HeesBlaizotBerges} as well as scalar theories coupled to a
fermionic field \cite{Reinosa:2005pj}, and more recently in the framework of QED
\cite{Reinosa:2006cm} in the covariant gauge. In this latter case, it is
important to emphasize that the renormalization procedure differs substantially
from the one in perturbation theory. The reason for this is that, for a given
loop truncation of the 2PI effective action and in contrast to what
happens in perturbation theory, the photon two- and four-point functions
develop longitudinal quantum and thermal corrections. Although these
contributions are formally of higher order than the order of the truncation,
they bring UV divergences which need to be removed before defining a continuum
limit. In Ref.~\cite{Reinosa:2006cm} a renormalization procedure involving a
new class of counterterms has been put forward which allows one to deal with this
new kind of UV divergences and thus opens the way to practical calculations.

% aspect of renormalization%]]]
% goal %[[[

In this letter, we apply these ideas in order to evaluate the pressure of QED
from the two-loop 2PI effective action. By solving the stationarity equations
for the propagators $\bar G$ and $\bar D$ and properly determining the
counterterms at the temperature of interest we calculate the pressure as given
in Eq.~\ref{eq:pressure}. This numerical procedure is repeated for several
gauge fixing parameters and for various renormalization scales in order to
explore how severe the problem of gauge dependence is in the context of
thermodynamic calculations within the 2PI framework.

% goal %]]]
% Introduction %]]]
% Lattice action %[[[

In this work, where we discuss gauge parameter dependence, it is essential
that the considered discretization respects gauge symmetry. In this way, the only source for gauge dependences is the particular truncation we use. For numerical purposes it is also convenient to use lattice rather than dimensional regularization. We thus consider QED on a hypercubic lattice of spacing $a$. We
denote by $N_\beta$ the number of points on the time direction and $N$ the
number of points on each of the spatial directions. The inverse temperature is
$\smash{\beta=N_\beta a}$ and the spatial volume $\smash{V=N^3a^3}$. We decompose the lattice
action in three pieces: $S=S_g+S_{gf}+S_f$. As gauge-field action, we consider
the non-compact action
\begin{equation}
S_g=\frac{1}{4}\,a^4\sum_x\sum_{\mu\nu} F_{\mu\nu}(x)F_{\mu\nu}(x)\,,
\end{equation}
where the field-strength tensor $\smash{F_{\mu\nu}(x)=\Delta_{\mu}^f
A_{\nu}(x)-\Delta_{\nu}^f A_{\mu}(x)}$ is expressed in terms of the forward
derivative\footnote{The notation $\hat\mu$ stands for the vector of length $a$ along the positive $\mu$ direction.}
$\smash{\Delta_{\mu}^fA_{\nu}(x)=a^{-1}\left[A_{\nu}(x+\hat\mu)-A_{\nu}(x)\right]}$.
We use a discretized covariant gauge fixing term
\begin{equation}
S_{gf}=\frac{1}{2\xi}\,a^4\sum_x\sum_{\mu\nu} \Delta_\mu^b
A_\mu(x)\,\Delta_\nu^b A_\nu(x) \,,
\end{equation}
given in terms of the backward derivative
$\smash{\Delta_{\mu}^bA_{\nu}(x)=a^{-1}[A_{\nu}(x)-A_{\nu}(x-\hat\mu)]}$ for
latter convenience. Finally, the fermionic action is taken to be the naive
chiral action
\begin{equation}
S_f=-\frac{1}{2a}\,a^4\sum_x\left[
\bar\psi(x+\hat\mu)\gamma_{\mu}U_{\mu}(x)\psi(x)-\bar\psi(x)\gamma_{\mu}U_{\mu}^+(x)\psi(x+\hat\mu)\right]\,,
\end{equation}
where $\smash{U_\mu(x)=\exp\left(iaeA_\mu(x)\right)}$ represents a link
variable. 

% Lattice action %]]]
% Lattice propagators & symmetries %[[[

Normally, the interacting two-point function $\bar D(x,y)$ or $\bar G(x,y)$ corresponds
to the correlator of two operators at $x$ and $y$. On the lattice however, where the fundamental objects are link variables, it is more convenient to introduce these two-point functions as
\begin{equation}\label{eq:prop_def}
\bar D(x,y)=\langle\psi(x)\bar\psi(y)\rangle_c \;\;\; {\rm and} \;\;\; \bar
G_{\mu\nu}(x,y)=\langle A_\mu(x)A_\nu(y-(1-\delta_{\mu\nu})\hat\nu)\rangle_c\,.
\end{equation}
This definition maintains 
the usual translation and reflection symmetries of $\bar G$ and $\bar D$,
as well as the identity $\smash{\bar G_{\mu\nu}(x,y)=\bar G_{\nu\mu}(x,y)}$.
Notice also that the discretization we consider here, respects the chiral symmetry of our massless fermion:
 $\smash{\bar D(x,y)=\sum_\mu \gamma_\mu \bar D_\mu(x,y)}$. 
We shall thus consider the 2PI effective action as a
functional of $D_\mu$ rather than a functional of $D$.

% Lattice propagators %]]]
% truncation %[[[

Since the pressure (\ref{eq:pressure}) cannot be determined exactly, we
consider the loop expansion of the 2PI
effective action as obtained from the Cornwall-Jackiw-Tomboulis formula
\cite{Cornwall:1974vz} (a trivial term stemming from the ghosts is included in our numerics but not written explicitly here):
\beq\label{eq:CJT}
\Gamma_{\rm 2PI}[D,G]=-N_f\,\Tr\! \left[\log
D^{-1}\!\!+D_0^{-1}D\right]+\frac{1}{2}\,\Tr\!\left[\log
G^{-1}\!\!+G_0^{-1}G\right]+\Gamma_{\rm int}[D,G]
\eeq
where we have defined $\smash{\Tr O\equiv a^4\sum_x\sum_iO_{ii}(x,x)=\beta
V\sum_i O_{ii}(x=0)}$ and we have included the possibility of an arbitrary
number of fermionic flavors $N_f$.\footnote{
The parameter $N_f$ will be used in what follows in order to eliminate the
doublers which appear as a result of discretizing the fermionic action. Since
our discretization generates $16$ fermion tastes (one pair in each direction)
we shall set $N_f$ to $1/16$. This is similar, for instance, to the fourth root
taken on the staggered fermion determinant in the context of lattice gauge
field theory \cite{Bernard:2006zw}.
}
The functional $\Gamma_{\rm int}[D,G]$ is given -- up to an overall sign -- by
all 0-leg 2PI diagrams that one can draw using the two-point functions $D$ and
$G$ and the tree level vertices generated by the lattice action. These arise
from the expansion of the link variable $U_\mu(x)$, and in turn of $S_f$, in
powers of $A_\mu(x)$. To make sure that the pressure we calculate is correct
up to $\mathcal{O}(e^3)$, we have to expand $U_\mu(x)$ to $\mathcal{O}(e^2)$. The vertex obtained from expanding $U_\mu(x)$ to $\mathcal{O}(e^3)$ brings no contribution to the pressure in the case of vanishing background fields, which we assume throughout this work.

Combining the $\mathcal{O}(e)$ and $\mathcal{O}(e^2)$
vertices into two-loop 2PI diagrams, performing the relevant traces and making use of the properties of $D$, we obtain the following contributions to the interacting part $\Gamma_{\rm int}$ of the 2PI effective action:
\begin{eqnarray}
\frac{1}{\beta V}\,\Gamma_{\rm int}^a & \!\!\!\!=\!\!\!\! & e^2N_f\,a^4\!\!
\sum_{x,\,\mu\ne \nu}
\!\!G_{\mu\nu}(x)\Big[D_\mu(x)D_\nu(x+\hat\mu+\hat\nu)+D_\nu(x)D_\mu(x+\hat\mu+\hat\nu)\nonumber\\
& & \hspace{3.0cm}
+\,D_\mu(x+\hat\nu)D_\nu(x+\hat\mu)+D_\nu(x+\hat\nu)D_\mu(x+\hat\mu)\Big]\nonumber\\\nonumber\\
& \!\!\!\!+\!\!\!\! & e^2N_f\,a^4\sum_{x,\,\mu}
G_{\mu\mu}(x)\Big[2D_\mu(x-\hat\mu)D_\mu(x+\hat\mu)+2D_\mu(x)D_\mu(x)\nonumber\\
& & \hspace{3.0cm}-\sum_\nu \left[D_\nu(x-\hat\mu)D_\nu(x+\hat\mu)+D_\nu(x)D_\nu(x)\right]\Big]
\label{eq:Gamma1}
\\
\frac{1}{\beta V}\Gamma_{\rm int}^b & \!\!\!\!=\!\!\!\! &
a\,e^2N_f\,\sum_\mu
G_{\mu\mu}(x=0)\Big[D_\mu(x=\hat\mu)-D_\mu(x=-\hat\mu)\Big].
\label{eq:Gamma2}
\end{eqnarray}

The contribution $\Gamma_{\rm int}^a$ is the usual fermion loop with a
somewhat peculiar photon line (\ref{eq:prop_def}). The lattice spacing
$a$ in $\Gamma_{\rm int}^b$ manifests that this diagram is a lattice artefact.
Both fermion loops are individually
quadratically divergent, they together make sure that at the lowest
{\em perturbative} level the photon receives no mass renormalization.

Together with a counterterm contribution $\delta\Gamma_{\rm int}$ (see below), the expressions~(\ref{eq:Gamma1}) and (\ref{eq:Gamma2}) provide the full $\mathcal{O}(e^2)$ interaction part of the 2PI effective action:
$\Gamma_{\rm int}=\Gamma_{\rm int}^a+\Gamma_{\rm int}^b+\delta\Gamma_{\rm int}$.
% truncation %]]]
% SD equatons %[[[
If we now introduce the self-energies
\begin{equation}
\bar\Sigma_\mu(x)=\bar
D_\mu^{-1}(x)-D_{0,\,\mu}^{-1}(x) \;\; {\rm and} \;\;\bar\Pi_{\mu\nu}(x)=G_{\mu\nu}^{-1}(x)-G_{0,\,\mu\nu}^{-1}(x)\,,\label{eq:1}
\end{equation}
and use the explicit formula (\ref{eq:CJT}) for the 2PI effective action, we
can write the stationarity equations defining the interacting two-point
functions $\bar D$ and $\bar G$ as
\begin{equation}
4N_f\,\bar \Sigma_\mu(x)=\frac{1}{a^4\beta V}
\frac{\partial\Gamma_{\rm int}}{\partial D_\mu(x)}\;\;\; {\rm and} \;\;\;
\bar \Pi_{\mu\nu}(x)=\frac{2}{a^4\beta V}
\frac{\partial\Gamma_{\rm int}}{\partial G_{\mu\nu}(x)}\,.
\label{eq:3}
\end{equation}
The interacting two-point functions $\bar D$ and $\bar G$ are thus obtained
after simultaneously solving Eqs.~(\ref{eq:1})-(\ref{eq:3}). As it can easily
be checked, in the two-loop approximation that we consider here,
Eqs.~(\ref{eq:3}) do not involve any
discretized integral in direct space. On the other hand, 
Eqs.~(\ref{eq:1}) can be conveniently solved in momentum
space.
% SD equatons %]]]
% Fourier space %[[[
To this order of the truncation we can thus completely avoid calculating loops
by simply Fourier transforming the propagators
back and forth in every step of the iterative procedure. 
We define the Fourier transforms of a generic fermionic ($D$) or
gauge ($G$) two-point function, respectively, as
\begin{equation}\label{eq:FT}
i^{-1}D_\mu(k)=a^4\sum_x \,e^{-ik\cdot x}\,D_\mu(x) \;\;\; {\rm and} \;\;\;
\alpha_{\mu\nu}^{-1}(k)G_{\mu\nu}(k)=a^4\sum_x\, e^{-ik\cdot x}\,G_{\mu\nu}(k)\,,
\end{equation}
where $\smash{\alpha_{\mu\nu}(k)=1}$ if $\smash{\mu=\nu}$ and
$\smash{\alpha_{\mu\nu}(k)=\exp(-ia(k_\mu+k_\nu)/2)}$ otherwise. This
particular definition of the Fourier transform of $G$ is connected to the fact
that the gauge field has to be thought as attached to the midpoints of the
links. Even this unusual variant of the fast Fourier transformation is
available as legacy code~\cite{FFTW}. Solving Eqs.~(\ref{eq:1}) in Fourier space, needs that we determine the Fourier transforms of the free inverse propagators. After inspection of the free (quadratic) contribution to $S$, one obtains 
\begin{equation}
D_{0,\,\mu}^{-1}(k)=-\bar k_\mu \;\;\; {\rm and} \;\;\; G^{-1}_{0,\,\mu\nu}(k)=\hat{k}^2\delta_{\mu\nu}-(1-\xi^{-1})\hat{k}_\mu\hat{k}_\nu\,,
\end{equation}
with the usual short-hand notations $\smash{\bar k_\mu a=\sin(k_\mu a)}$ and
$\smash{\hat k_\mu a=2\sin(k_\mu a/2)}$.
% Fourier space %]]]
% counterterms %[[[

Eqs.~(\ref{eq:1})-(\ref{eq:3}) can be solved for any non-vanishing lattice spacing $a$, leading to perfectly finite two-point functions $\bar D$ and $\bar G$. In order to define a proper continuum limit of the latter, as $\smash{a\rightarrow 0}$, one needs however to absorb UV divergences. Renormalization of Eqs.~(\ref{eq:1}) and (\ref{eq:3}) was considered in Ref.~\cite{Reinosa:2006cm} in the context of dimensional
regularization and at zero temperature. There, renormalization was achieved by
adding a contribution $\delta\Gamma_{\rm int}$ to the functional $\Gamma_{\rm
int}$. This contribution carries the counterterms needed for renormalization.
In extending this result to lattice regularization, one has to pay attention to
the presence of new vertices originating from the expansion of the link
variable $U_\mu(x)$ in powers of the field $A_\mu(x)$ (see above). In our
present calculation, in addition to the usual vertex coupling $A$ to $\bar\psi$
and $\psi$ (which leads to $\Gamma_{\rm int}^a$), there is a new vertex coupling $A^2$ to $\bar\psi$ and $\psi$ (which leads to $\Gamma_{\rm int}^b$). This new vertex brings an extra factor of $a$,
which is such that the superficial degree of
divergence of a given diagram is the same as in dimensional regularization.\footnote{Namely $\smash{\delta=4-E_A-(3/2)E_\psi}$, where $E_A$ and $E_\psi$
respectively denote the number of external photon and fermion legs of the
diagram at hand.} It follows that we can here apply the same type of analysis of UV
divergences as the one used in Ref.~\cite{Reinosa:2006cm}. 

At two-loop order, the shift $\delta\Gamma_{\rm int}$ is given in
lattice regularization by
\begin{eqnarray}
\delta\Gamma_{\rm int} & \!\!\!=\!\!\! & \frac{\delta
g_1}{8}\frac{1}{\beta
V}\sum_{k,\mu}G_{\mu\mu}(k)\sum_{q,\nu}G_{\nu\nu}(q)+\frac{\delta
g_2}{4}\frac{1}{\beta V}\sum_{\mu\nu}\sum_k G_{\mu\nu}(k)\sum_q
G_{\mu\nu}(q)\nonumber\\
& \!\!\!+\!\!\! & \frac{1}{2}\sum_k\sum_{\mu\nu}G_{\mu\nu}(k)\Big[\delta
Z_3\hat k^2\delta_{\mu\nu}-(\delta Z_3-\delta\lambda)\hat k_\mu\hat
k_\nu+\delta M^2\delta_{\mu\nu}\Big]\nonumber\\
& \!\!\!-\!\!\! & 4N_f\,\delta Z_2\sum_{k,\mu} \bar k_\mu D_\mu(k)\,.
\end{eqnarray}
It leads to additional contributions at the level of the self-energies, in particular  a longitudinal
wave function renormalization ($\delta\lambda$) as well as a photon mass
counterterm ($\delta M^2$).

The counterterms $\delta g_1$ and $\delta g_2$ allow to remove subdivergences
hidden in Eqs.~(\ref{eq:3}) and involving four photon legs (see below). After
these have been removed, there only remain temperature independent overall
divergences that need to be absorbed in the counterterms $\delta Z_2$,
$\delta Z_3$, $\delta \lambda$ and $\delta M^2$. 
% counterterms %]]]
% structure of divergencies %[[[
Although the exact O(4) symmetry is broken on the lattice, the tensor structure
of the self energies at a fixed scale $k$ is restored in the
continuum limit of an isotropic lattice theory. This allows us to use the
renormalization conditions introduced in the context of the continuum theory
\cite{reisz}.
In particular, one can show that the overall divergences have the structure
\begin{equation}
\bar\Sigma_\mu^{\rm div}(k)=-\sigma {\bar k}_\mu \;\; 
{\rm and} \;\;
\bar\Pi_{\mu\nu}^{\rm div}(k)=\pi_M\delta_{\mu\nu}+
\pi_T(\delta_{\mu\nu}{\hat k}^2-{\hat k}_\mu {\hat k}_\nu)
+\pi_L k_\mu k_\nu 
\end{equation}
where $\sigma$, $\pi_T$, $\pi_L$ and $\pi_M$ represent
quantities which diverge as $\smash{a\rightarrow 0}$ (quadratically for $\pi_M$ and
logarithmically for the rest of them). 
Comparing these expressions to those for the counterterms, we find that
all divergences can be absorbed by setting
\begin{equation}\label{eq:absorb}
\delta Z_2=-\sigma\,,\;\;\delta Z_3=-\pi_T\,,\;\;\delta\lambda=-\pi_L \;\; {\rm and} \;\; \delta M^2=-\pi_M\,.
\end{equation}

% structure of divergencies %]]]
% renormalization conditions %[[[

The set of Eqs.~(\ref{eq:absorb}) does not fix the finite parts of the counterterms. In
order to do so, we fix $\delta Z_2$, $\delta Z_3$, $\delta\lambda$ and $\delta
M^2$ through the renormalization conditions
\begin{equation}\label{eq:renorm}
\left.\frac{\partial\bar\Sigma_3^\star}{\partial \bar k_3}\right|_{k^\star}=0\,,\;\;\left.\frac{\partial\bar\Pi_{22}^\star}{\partial \hat k^2_3}\right|_{k^\star}=0\,,\;\;\left.\frac{\partial\bar\Pi_{33}^\star}{\partial \hat k_3^2}\right|_{k^\star}=0\,,\;\; {\rm and} \;\; \bar\Pi_{33}^\star|_{k^\star}=0\,,
\end{equation}
where $\smash{k^\star=(0,0,\mu,0)}$ and $\mu$ denotes our renormalization scale. The
star on the self-energies means that these are considered at a reference
temperature $T^\star$. The first two renormalization conditions are similar to
those which are used in perturbation theory and completely determine the
counterterms $\delta Z_2$ and $\delta Z_3$. 
In perturbation theory, where the (lattice) Ward identity for $\bar\Pi(k)$
\begin{equation}
0=\sum_\mu \hat k_\mu\bar\Pi_{\mu\nu}(k)
\end{equation}
prevents the appearance of longitudinal corrections to the self energy, the third and fourth conditions in
Eq.~(\ref{eq:renorm}) are trivially satisfied.  In our case, however, we need to
fix two counterterms ($\delta\lambda$ and $\delta M^2$) that cancel UV
divergences of $\mathcal{O}(e^4)$.  A natural way to fix these is to impose
the Ward identity on $\bar\Pi$ at the renormalization point
$\smash{k^\star=(0,0,\mu,0)}$ and for a given temperature $T^\star$.
We do so at $k^\star$ and in a small neighborhood of $k^\star$. In this way, we obtain the third
and fourth renormalization conditions in Eq.~(\ref{eq:renorm}). The arbitrariness
of this condition introduces an ambiguity of order $\mathcal{O}(e^4)$.

% renormalization conditions %]]]
% Bethe-Salpeter %[[[

As already discussed in Ref.~\cite{Reinosa:2006cm}, when renormalizing the
two-point function $\bar G$, one has not only to pay attention to longitudinal
overall divergences but also to longitudinal subdivergences which involve
four-photon legs. Again, if no truncation is considered, these subdivergences
automatically cancel since they reproduce the exact four-photon function which
is transverse.  However, for a given truncation of the 2PI effective action,
this cancellation of divergences is only true up to the order of the
truncation. Above, new divergences appear which need to be absorbed by means of
the counterterms $\delta g_1$ and $\delta g_2$. The particular structure of
these divergences has been worked out in Ref.~\cite{Reinosa:2006cm} for the
case of dimensional regularization. The result is that, in order to absorb the
four-photon divergences, one needs to impose,
at the renormalization point, the transversality of a four-point function
defined by means of a set of Bethe-Salpeter equations. Here, we extend this
result to the case of lattice regularization. 

The Bethe-Salpeter equations can be written as a closed set of equations for a four-point function $\bar V_{\mu\nu,\sigma\rho}(p,k)$ involving four photon legs and a four-point function $\bar W_{ij,\sigma\rho}(p,k)$ involving two photon and two fermion legs \cite{Reinosa:2006cm}.  Similarly to what we did with the propagator $\bar D$, we turn the Dirac indices ${i,j}$ into one Lorentz index $\mu$: 
$\smash{\sum_\mu \bar W_{\mu,\sigma\rho}{\gamma_\mu}_{ij}=\bar W_{ij,\sigma\rho}}$.
Given that $\smash{k^\star=(0,0,\mu,0)}$, the renormalization conditions fixing $\delta g_1$ and $\delta g_2$, as given in Ref.~\cite{Reinosa:2006cm}, read
% V renormalization conditions [[[
\begin{equation} 
\bar V_{2233}(k^\star,k^\star)=0\;\; {\rm and} \;\;
\bar V_{3333}(k^\star,k^\star)=0\,.
\label{eq:Vcond}
\end{equation} %]]]

In order to impose these renormalization conditions, we do not need to solve
the set of Bethe-Salpeter equations for arbitrary values of the momenta and arbitrary
configurations of Lorentz indices. Indeed, the set of equations remains closed
if we fix one of the momenta to $\smash{k=k^\star}$ and two of the Lorentz
indices to $\smash{\sigma=\rho=3}$. We shall thus consider equations for $\bar
V_{\mu\nu}(p)=\bar V_{\mu\nu33}(p,k^\star)$ and $\bar W_{\alpha}(p)=\bar
W_{\alpha,33}(p,k^\star)$. Introducing the notations
% A,Vbar,Wbar notations [[[
\begin{eqnarray} 
A_{\sigma\rho}(p)&\!\!\!=\!\!\!&\delta(p-k^\star)\delta_{\sigma3}\delta_{\rho3}\,,\\
V_{\mu\nu}(p)&\!\!\!=\!\!\!&\bar G_{\mu\alpha}(p)\,\bar V_{\alpha\beta}(p)\,\bar G_{\beta\nu}(p)\,,\\
W_\mu(p)&\!\!\!=\!\!\!&-2\,\bar D_\mu(p)\sum_\rho\bar W_\rho(p)\bar D_\rho(p)
+\bar W_\mu(p)\sum_\rho \bar D_\rho(p) \bar D_\rho(p)\,,
\end{eqnarray} %]]]
we may write the corresponding set of Bethe-Salpeter equation as
% bethe salpter equation [[[
\begin{eqnarray} 
\bar V_{\mu\nu}(p)&\!\!\!=\!\!\!&-\frac{\delta_{\mu\nu}}{2}
\frac{\delta g_1}{\beta V}\sum_{q,\rho}\left[V_{\rho\rho}(q)-2A_{\rho\rho}(q)\right]
-\frac{\delta g_2}{\beta V}\sum_{q}\left[V_{\mu\nu}(q)-2A_{\mu\nu}(q)\right]\nonumber\\
&& -\sum_{q,\rho}\frac{\partial \Pi_{\mu\nu}(p)}{\partial D_\rho(q)}W_\rho(q)\,,\\
\bar W_{\mu}(p)&\!\!\!=\!\!\!&
-\sum_{q,\rho\sigma}\frac{\partial\Sigma_{\mu}(k)}{\partial G_{\rho\sigma}(q)}
\left[V_{\rho\sigma}(q)-2A_{\rho\sigma}(q)\right]
-\sum_{q,\rho}\frac{\partial\Sigma_{\mu}(p)}{\partial D_{\rho}(q)}
W_{\rho}(q)
\end{eqnarray} %]]]

We solved this pair of equations iteratively by adjusting $\delta g_1$ and
$\delta g_2$ after each step so that Eq.~(\ref{eq:Vcond}) is always fulfilled.
As expected, the numerical values of these counterterms scale as
$\sim e^4\log(a)$.
% Bethe-Salpeter %]]]
%  the algorithm %[[[
Once the counterterms have been fixed according to the renormalization
conditions (\ref{eq:renorm}) and (\ref{eq:Vcond}), we can solve for the physical
two-point functions $\bar D$ and $\bar G$ which admit a proper continuum limit.
Plugging this values into the CJT formula (\ref{eq:CJT}) truncated at two-loop
order gives us a non-perturbative approximation to the QED pressure, compatible
with perturbation theory up to order $\mathcal{O}(e^3)$. Notice that, even
with all our counterterms, there is a quartic
divergence remaining in the pressure. This divergence is temperature
independent and can be removed by a `cosmological constant' renormalization.
The renormalization condition is usually given by the requirement of zero
vacuum  pressure. Here we do not renormalize or evaluate the model at zero
temperature. We determine the counterterms in the equations of motion at
$T^\star$. Then, using these counterterms we evaluate the pressure at $T^\star$
and $T^\star/2$. Assuming a $\sim T^4$ scaling with the temperature, we
determine the pressure as the difference of the divergent pressure values as
obtained from the formula of the effective action, divided by
$(15/16){(T^\star})^4$. The assumed scaling of temperature is broken due to the presence of the renormalization scale. This effect introduces an error of $\mathcal{O}(e^4)$ which is
above the actual accuracy of our calculation. 

% plot pvse %[[[
\begin{figure}[ht]%[[[
\begin{center}
\includegraphics[width=10cm]{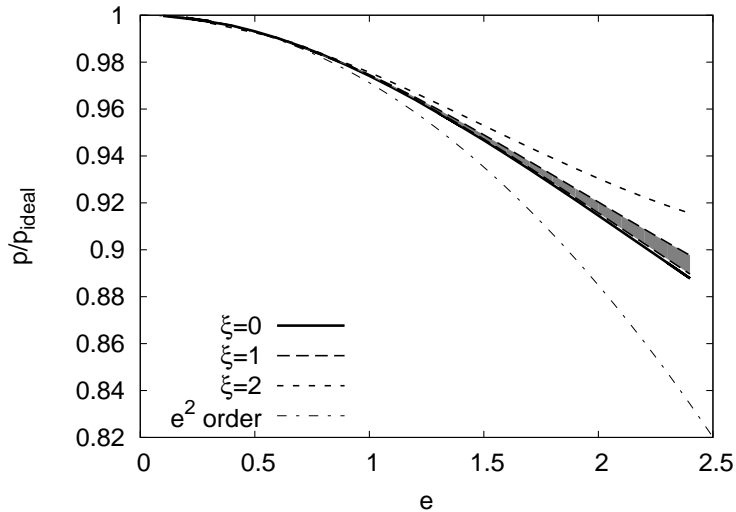}
\caption{Two-loop QED pressure as a function of the coupling $e$ and for
different values of the gauge-fixing parameter $\xi$. The plain line
corresponds to $\xi=0$ (Landau gauge), long-dashed lines to $\xi=1$ (Feynman
gauge) and short-dashed lines to $\xi=2$. The sensitivity with respect to the
renormalization scale $\mu$ is illustrated in the case of the Feynman gauge. We
also plot the perturbative $\mathcal{O}(e^2)$ result for comparison.}
\label{fig:pvse}
\end{center}
\end{figure}%]]]
% plot pvse %]]]

In order to improve numerical stability, we took into account the following points. Calculating
the pressure difference involves the subtraction of two quartically divergent
contributions. Instead, we carried out the spatial part of the trace in
$\Gamma_{\rm 2PI}$ {\em after} performing the subtraction. An other important alteration to the equations above was the exclusion of the
spatially homogeneous lattice mode on the level of the 2PI effective action.
This was necessary to avoid instabilities as $e\to0$, since the finite
photon mass contribution behaves as $\sim e^4$.

% the algorithm %]]]
% plot pvslrs %[[[
\begin{figure}[ht]
\begin{center}
\centerline{
\includegraphics[width=7cm]{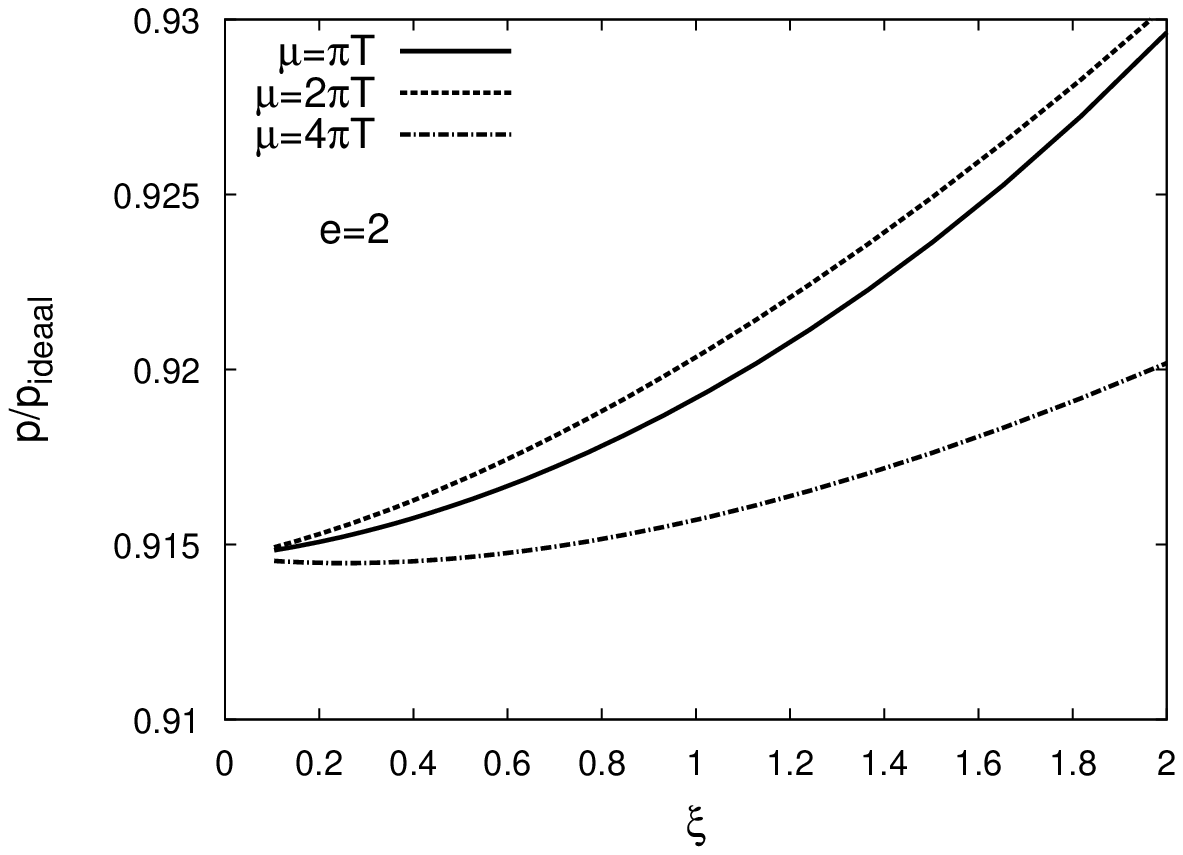}
\includegraphics[width=7cm]{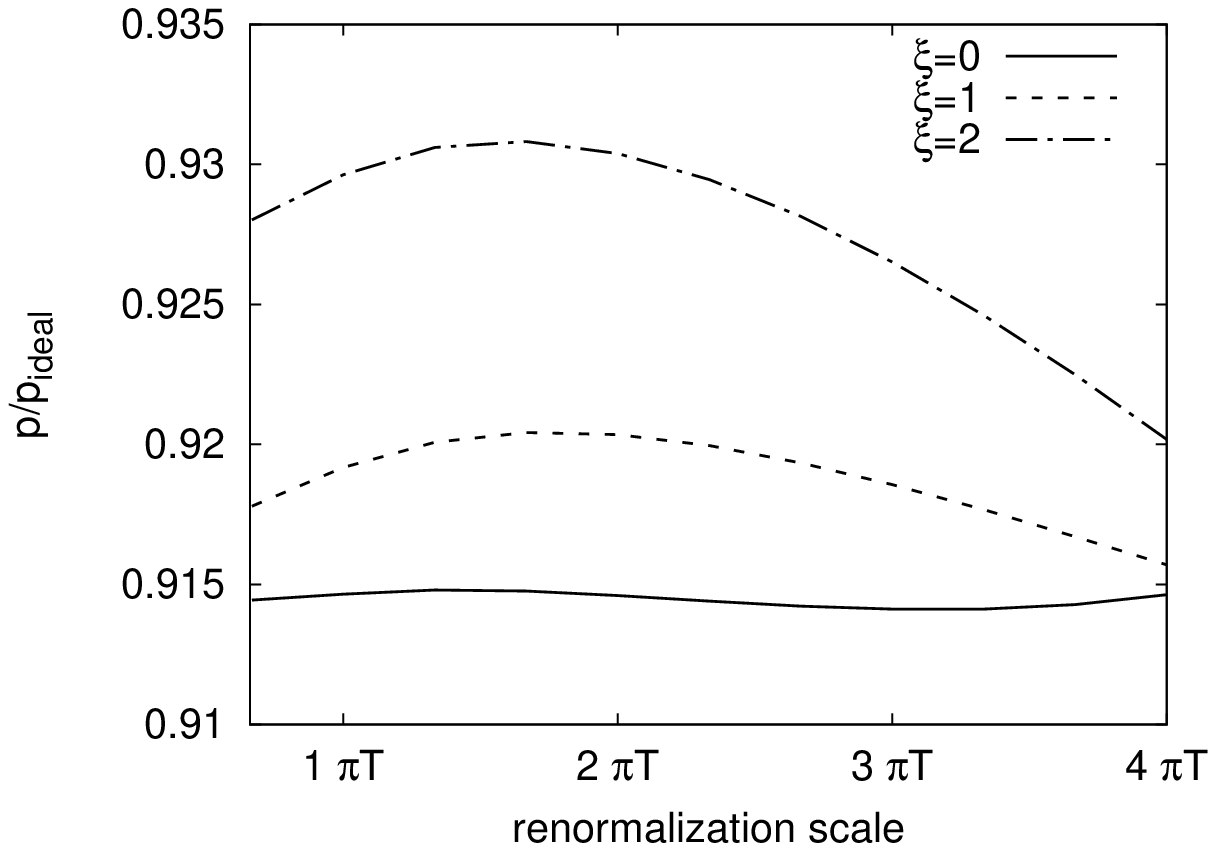}
}
\caption{Renormalization scale dependence of the two-loop pressure.}
\label{fig:pvslrs}
\end{center}
\end{figure}
% plot pvslrs %]]]
% pressure vs coupling %[[[
In Fig.~\ref{fig:pvse} we plot the QED pressure in the two-loop 2PI
approximation, for a wide range of coupling values ($\smash{0\leq e \leq 2.4}$)
and for various values of the gauge-fixing parameter. As discussed in
Ref.~\cite{Arrizabalaga:2002hn} the higher the gauge-fixing parameter is, the less
convergent the 2PI loop expansion becomes. It is thus meaningless to consider
our calculation for too high values of $\xi$ and, as suggested in
Ref.~\cite{Arrizabalaga:2002hn}, we restricted our calculations to values of the
gauge-fixing parameter ranging from $\smash{\xi=0}$ (Landau gauge) to
$\smash{\xi=2}$. For small values of the coupling, our results are almost
insensitive to the gauge fixing parameter and nicely reproduce the perturbative
result to order $\mathcal{O}(e^2)$. This comes as no surprise since the
two-loop 2PI approximation contains all diagrams contributing to order
$\mathcal{O}(e^3)$.\footnote{
Our calculation reproduces the $\mathcal{O}(e^3)$ result in the infinite
volume limit only.
} Numerically we find a good agreement with perturbation theory up
to $e\sim 1$ which is precisely where the perturbative expansion usually breaks
down. 

For large values of the coupling, our calculation becomes a priori sensitive to
two types of uncertainties. First of all, renormalization is done by imposing
renormalization conditions at a certain momentum $\smash{k^\star=(0,0,\mu,0)}$
which introduces an artificial dependence on the scale $\mu$. Moreover the
truncation of the 2PI effective action introduces gauge parameter dependences
starting at order $\mathcal{O}(e^4)$. These two types of uncertainties can be
taken as a way to estimate the error of the calculation. 

% pressure vs coupling %]]]
% scale dependence %[[[
The dependence with respect to the renormalization scale $\mu$ is illustrated
in Fig.~\ref{fig:pvse} for the case $\smash{\xi=1}$ (Feynman gauge) where $\mu$
is varied in the interval $\smash{\pi T\leq \mu \leq 4\pi T}$ as it is usually
done in calculations at finite temperature. A study of the $\mu$-dependence as
the gauge-fixing parameter is varied and for a given value of the coupling is
depicted in Fig.~\ref{fig:pvslrs}. Notice that, at fixed gauge-fixing parameter
$\xi$, the $\mu$-dependence is not monotonous. However $\mu=2\pi T$ roughly
represents the value at which the pressure reaches it maximum value, in this range. 
We notice that the uncertainty due to scale dependence is not
particularly severe, which indicates the good convergence behavior of the
2PI approach.  Moreover this uncertainty is $\sim 1{\%}$
for $\smash{\xi=2}$ and decreases considerably down to
its minimum value reached for $\smash{\xi=0}$, which makes the Landau gauge a
particularly interesting choice among all possible gauges. We also notice that,
in general, choosing a higher renormalization scale flattens the gauge
dependence towards the Landau gauge value.

\begin{figure}[thb]
\begin{center}
\includegraphics[width=10cm]{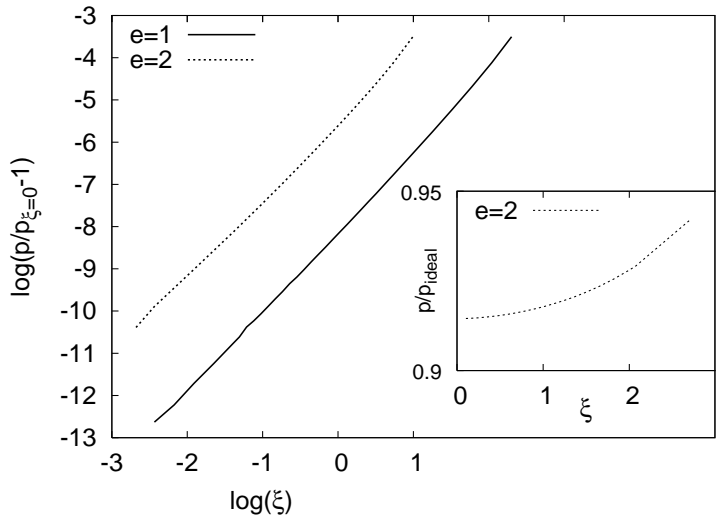}
\caption{Gauge-fixing parameter dependence of the two-loop pressure.}
\label{fig:pvsl}
\end{center}
\end{figure}

% scale dependence %]]]
% gauge fixing dependence %[[[
The second source of uncertainties is gauge dependence.
As already mentioned a calculation for high values of the gauge fixing
parameter makes little sense. In the considered range of gauge parameter values, the error due to gauge dependence is of the order of or less than $1-1.5\%$. The Landau gauge plays again a special role since it corresponds to the value of $\xi$ for which the pressure is the less sensitive to gauge parameter dependence. Indeed, independently of the value of the
coupling, one has $p_\xi-p_{\xi=0}\sim \xi^2$ as $\xi\rightarrow 0$, as it is
clear on the logarithmic plot of Fig.~\ref{fig:pvsl}.

% gauge dependence %]]]
% plot pvsl %[[[

% plot pvsl %]]]
% Conclusions %[[[

In conclusion, our calculation shows, in covariant gauge, a relatively small error coming
from gauge parameter dependence. The parametric suppression of the gauge parameter
dependence has already been shown in Ref.~\cite{Arrizabalaga:2002hn}. We have now established that the so far unknown coefficients of this parametric
dependence do not spoil this behavior. The gauge dependence can also
be regarded as a bonus feature, which opens a way to error estimates
without the need for considering higher order diagrams. We think,
that the 2PI effective action can be regarded as an efficient resummation
technique for gauge theories, where the actual choice of gauge fixing
has an impact on the quality of the resummation. As for the particular calculation presented here,
the Landau gauge is the preferred choice. If this result persists in QCD,
it could serve as a justification for the exclusive use of Landau gauge in
the QCD Schwinger-Dyson equations \cite{landauQCD}.

% Conclusions %]]]
% acknowledgements %[[[

\vspace{1mm}
{\bf Acknowledgments:}
UR acknowledges support from the Alexander von Humboldt foundation. The authors
thank J.~Berges and J.~Serreau 
for inspiring discussions and for collaboration on related work.
%The computational resources have been provided by the University of Sussex.
\vspace{-4.5mm}
% acknowledgements %]]]
% references %[[[


\begin{thebibliography}{10}

\bibitem{resum}
%\bibitem{Kraemmer:1994az}
  U.~Kraemmer, A.~K.~Rebhan and H.~Schulz,
  %``Resummations In Hot Scalar Electrodynamics,''
  Annals Phys.\  {\bf 238} (1995) 286;
  %%CITATION = APNYA,238,286;%%
%\bibitem{Blaizot:2001ev2003iq}
  J.~P.~Blaizot, E.~Iancu and A.~Rebhan,
  %``The thermodynamics of the quark-gluon plasma: Self-consistent  resummations
  %versus lattice data,''
  Nucl.\ Phys.\  A {\bf 698} (2002) 404;
  %%CITATION = NUPHA,A698,404;%
% \bibitem{Blaizot:2003iq}
 % J.~P.~Blaizot, E.~Iancu and A.~Rebhan,
  %``On the apparent convergence of perturbative QCD at high temperature,''
  Phys.\ Rev.\  D {\bf 68} (2003) 025011;
  %%CITATION = PHRVA,D68,025011;%%
%\cite{Andersen:2004fp}
%\bibitem{Andersen:2004fp}
  J.~O.~Andersen and M.~Strickland,
  %``Resummation in hot field theories,''
  Annals Phys.\  {\bf 317} (2005) 281.
  %%CITATION = APNYA,317,281;%%


\bibitem{Luttinger:1960ua:andothers}
J.~M. Luttinger and J.~C. Ward, Phys. Rev. {\bf 118} (1960) 1417;
%\bibitem{Baym:1962sx}
G. Baym, Phys. Rev. {\bf 127} (1962) 1391; 
% \bibitem{DeDom:1964vz}
C. De Dominicis and P.~C. Martin, J .Math. Phys. {\bf 5} (1964) 14.

\bibitem{Cornwall:1974vz}
J.~M. Cornwall, R. Jackiw, and E. Tomboulis, Phys. Rev. D {\bf 10} (1974) 2428.

\bibitem{Berges:2004yj}
  J.~Berges,
  %``Introduction to nonequilibrium quantum field theory,''
  AIP Conf.\ Proc.\  {\bf 739} (2005) 3.
  %%CITATION = APCPC,739,3;

%\cite{Berges:2004hn}
\bibitem{Berges:2004hn}
  J.~Berges, S.~Borsanyi, U.~Reinosa and J.~Serreau,
  %``Renormalized thermodynamics from the 2PI effective action,''
  Phys.\ Rev.\  D {\bf 71} (2005) 105004.
  %%CITATION = PHRVA,D71,105004;%%


\bibitem{LeBellac}
M.~Le~Bellac, {\it Thermal Field Theory}, Cambridge University Press (1996).

  
\bibitem{Reinosa:2007vi}
%\bibitem{Reinosa:2004bn}
  U.~Reinosa,
  %``Renormalization and gauge symmetry for 2PI effective actions,''
  arXiv:hep-ph/0411255;
  %%CITATION = HEP-PH/0411255;%%
  U.~Reinosa and J.~Serreau,
  %``Ward Identities for the 2PI effective action in QED,''
  arXiv:0708.0971 [hep-th].
  %%CITATION = ARXIV:0708.0971;%%

\bibitem{Arrizabalaga:2002hn}
A. Arrizabalaga and J. Smit, Phys. Rev. D {\bf 66} (2002) 065014.


\bibitem{Blaizot:1999ip}
J.-P.~Blaizot, E.~Iancu, and A.~Rebhan, Phys. Rev. Lett. {\bf 83} (1999) 2906;
Phys. Lett. B {\bf 470} (1999) 181;
Phys. Rev. D {\bf 63} (2001) 065003.

\bibitem{Leupold:2006bp}
S.~Leupold, Phys. Lett. B {\bf 646} (2006) 155.

%\cite{Andersen:2004re}
\bibitem{Andersen:2004re}
  J.~O.~Andersen and M.~Strickland,
  %``Three-loop Phi-derivable approximation in QED,''
  Phys.\ Rev.\  D {\bf 71} (2005) 025011.
  %%CITATION = PHRVA,D71,025011;%%

\bibitem{Blaizot:2005wr}
J.-P.~Blaizot, A.~Ipp, A.~Rebhan and U.~Reinosa, Phys. Rev.  D {\bf 72} (2005) 125005.




\bibitem{HeesBlaizotBerges}
H.~van Hees and J.~Knoll, Phys.\ Rev.\ D {\bf 65}, 025010 (2002) 025010; Phys.\ Rev.\ D {\bf 65} (2002) 105005;
J.~P.~Blaizot, E.~Iancu and U.~Reinosa, Phys.\ Lett.\ B {\bf 568} (2003) 160; Nucl.\ Phys.\ A {\bf 736} (2002) 149;
J. Berges, S. Bors\'anyi, U. Reinosa, and J. Serreau, Annals Phys. {\bf 320} (2005) 344.

\bibitem{Reinosa:2005pj}
U.~Reinosa, Nucl. Phys. {\bf A772} (2006) 138.

\bibitem{Reinosa:2006cm}
U.~Reinosa and J.~Serreau, JHEP {\bf 07} (2006) 028.

\bibitem{Bernard:2006zw}
  C.~Bernard,
  %``Staggered chiral perturbation theory and the fourth-root trick,''
  Phys.\ Rev.\  D {\bf 73} (2006) 114503.
  %%CITATION = PHRVA,D73,114503;%%

\bibitem{FFTW}
 M.~Frigo and S.~G.~Johnson,
% "The Design and Implementation of FFTW3,"
  Proceedings of the IEEE 93 (2) (2005) 216.

%\bibitem{latticeO4}
%%\cite{Kogut:1981ny}
%%\bibitem{Kogut:1981ny}
%  J.~B.~Kogut, D.~K.~Sinclair, R.~B.~Pearson, J.~L.~Richardson and
%J.~Shigemitsu,
%  %``The Fluctuating String Of Lattice Gauge Theory: The Heavy Quark Potential,
%  %The Restoration Of Rotational Symmetry  And Roughening,''
%  Phys.\ Rev.\  D {\bf 23} (1981) 2945;
%  %%CITATION = PHRVA,D23,2945;%%
%%\cite{Lang:1982tj}
%%\bibitem{Lang:1982tj}
%  C.~B.~Lang and C.~Rebbi,
%  %``Potential And Restoration Of Rotational Symmetry In SU(2) Lattice Gauge
%  %Theory,''
%  Phys.\ Lett.\  B {\bf 115} (1982) 137.
%  %%CITATION = PHLTA,B115,137;%%

\bibitem{reisz}
  T.~Reisz,
  %``RENORMALIZATION OF FEYNMAN INTEGRALS ON THE LATTICE,''
  Commun.\ Math.\ Phys.\  {\bf 117} (1988) 79;
  %%CITATION = CMPHA,117,79;%%
  T.~Reisz,
  %``LATTICE GAUGE THEORY: RENORMALIZATION TO ALL ORDERS IN THE LOOP
  %EXPANSION,''
  Nucl.\ Phys.\  B {\bf 318} (1989) 417.
  %%CITATION = NUPHA,B318,417;%%


\bibitem{landauQCD}
%\cite{vonSmekal:1997is}
%\bibitem{vonSmekal:1997is}
  L.~von Smekal, R.~Alkofer and A.~Hauck,
  %``The infrared behavior of gluon and ghost propagators in Landau gauge
  %QCD,''
  Phys.\ Rev.\ Lett.\  {\bf 79} (1997) 3591;
  %%CITATION = PRLTA,79,3591;%%
%\cite{Pawlowski:2003hq}
%\bibitem{Pawlowski:2003hq}
  J.~M.~Pawlowski, D.~F.~Litim, S.~Nedelko and L.~von Smekal,
  %``Infrared behaviour and fixed points in Landau gauge QCD,''
  Phys.\ Rev.\ Lett.\  {\bf 93} (2004) 152002.
  %%CITATION = PRLTA,93,152002;%%
     
\end{thebibliography}
\end{document}